\documentclass[
aps,
prb,
twocolumn,
superscriptaddress,
showpacs,
english,
floatfix]
{revtex4-1}

\usepackage{float}
\usepackage{subcaption}
\usepackage[T1]{fontenc}
\usepackage[utf8]{inputenc}
\usepackage{babel}
\usepackage{amsmath}
\usepackage{amssymb}
\usepackage{wasysym}
\usepackage{graphicx}
\usepackage{multirow}
\usepackage{gensymb}
\usepackage[dvipsnames]{xcolor}
\usepackage{pifont}
\usepackage{microtype}
\usepackage{xcolor}
\usepackage{soul}
\usepackage[version=3]{mhchem} 
\usepackage{chemformula}[2014/04/08] %nice x's
\usepackage{booktabs}    % cmdirule
\usepackage{dcolumn}     % Align table columns on decimal point
\usepackage{bm}          % bold math
\usepackage{url}

\usepackage[linktocpage=true,
  colorlinks=true, 
  pdfborder={0 0 0},
  linkcolor=blue,
  citecolor=red,
  filecolor=yellow,
  urlcolor=blue,
  bookmarks,
  pdfauthor={},
]{hyperref}

\usepackage{cleveref}

\newcommand{\tc}{T$_{\text{c}}$}

\newcommand{\cubic}{$Im\overline{3}m$}

\newcommand{\figref}[1]{Fig.~\ref{#1}}

\newcommand{\Riken}{RIKEN Center for Emergent Matter Science, 2-1 Hirosawa, Wako, 351-0198, Japan}
\newcommand{\UniRoma}{Dipartimento di Fisica, Universit\`a di Roma La Sapienza, Piazzale Aldo Moro 5, I-00185 Roma, Italy}
\newcommand{\UniBasel}{Department\ of\ Physics,\ University\ of\ Basel,\ Klingelbergstrasse\ 82,\ CH-4056\ Basel,\ Switzerland}
\newcommand{\CSCS}{CSCS, Swiss National Supercomputing Centre, Via Trevano 131, CH-6900 Lugano, Switzerland}

\begin{document}

%\title{Missing theoretical evidence for the carbonaceous sulfur hydride room temperature superconductor} 

\title{Missing theoretical evidence for conventional room temperature superconductivity in low enthalpy structures of carbonaceous sulfur hydrides}

% Jose's titles: 
% Enthalpy landscape of the C-S-H ternary at high pressure; missing of a stable phase for the room-temperature superconductor  
% Missing a stable phase for the room temperature superconductor at high pressure in C-S-H polymorphs? 

\author{Moritz Gubler}          \affiliation{\UniBasel}
\author{Jos\'e A. Flores-Livas} \affiliation{\UniRoma} \affiliation{\Riken} 
\author{Anton Kozhevnikov}      \affiliation{\CSCS}
\author{Stefan Goedecker}       \affiliation{\UniBasel}

\begin{abstract}
To elucidate the geometric structure of the putative room temperature superconductor, carbonaceous sulfur hydride, at high pressure, we present the results of an extensive computational structure search of bulk C-S-H at 250\,gigapascals. 
Using the minima hopping structure prediction method coupled to the GPU accelerated Sirius library, more than 17,000 local minima with different stochiometries in large simulation cells were investigated. 
Only 24 stochiometries are favourable against elemental decomposition, all of them are carbon doped \ce{H3S} crystals.
The absence of van Hove singularities or similar peaks in the electronic density of states of more than 3.000 candidate phases rules out conventional superconductivity in C-S-H at room-temperature. 
\end{abstract}

\maketitle

%=============================================
\section{Introduction}\label{Sec:Int}
%=============================================

In late 2020 the group of Dias published evidence for the first room-temperature superconductor (RTS) at 289\,K and 267\,GPa~\cite{room_temp}. According to the authors, the superconductive material is carbonaceous sulfur hydride (CSH) produced by photochemical synthesis at high pressure. In their work, the authors initially mixed carbon and sulfur at a 1:1 molar ratio; then, the sample was put into gaseous hydrogen, pressurized in a diamond anvil cell and exposed to 532\,nm laser light, a wavelength known to break sulfur bonds. \textit{Unfortunately, despite the relative simplicity of the recipe, no other researchers were able to reproduce this synthesis pathway and the exact stochiometry and crystalline structure remain unknown.}

More recently, Lamichhane et al.~\cite{eos} examined the X-ray diffraction patterns for samples of CSH polymorphs and determined their equation of state. They pointed out that the structure could have the composition \ce{[(CH4)2]_x [(H2S)2H2]_y} where $y$ is much bigger than $x$. However, measurements of resistivity or magnetic response were not carried out to corroborate the superconducting state. Experimentally, the fact that hydrogen is involved with high pressure challenges the accurate determination of the possible crystalline structures. 

On the theoretical side, Sun et al.~\cite{100gpa_search} and Cui et al.~\cite{csh7_search} published the results of structure searches in the C-S-H ternary before the publication of the results of Dias's group. These theoretical studies focused on pressures of 100 and 200\,GPa, lower than the reported pressure for the RTS (close to 270\,GPa). However, extrapolating the results from lower to high pressures is unreliable since the chemistry changes drastically with every GPa gained. Nevertheless, not a single candidate was found to be thermodynamically stable among all the structures explored in these works.

A fresher look recently came from Wang and coworkers~\cite{Wang_absence_PRB-2021}, who studied the 250\,GPa domain, and ruled out doping as a possible explanation of the RTS. However, while their crystalline search was thorough, it focused on small doping regions near the known \ce{H3S} superconductor.

Moreover, a debate is rising in the literature concerning the uncommon superconducting features of the CSH compound~\cite{hirsch2020absence,hirsch2021absence,Hirsch_PRB_2021,talantsev2021electron,dogan2021anomalous}, namely a sharp drop of electric resistivity at \tc\ and an unusual dependence on magnetic fields~\cite{Matters_arising}. 

In this work, we studied the enthalpy landscape of C-S-H at two different pressures thoroughly. Thanks to the use of Graphical Processing Units (GPU), we could extend our searches to larger crystalline cells using highly accurate density functional theory (DFT). As a result, our search differs substantially from previous works in extent and depth. We organise the rest of the paper as follows: Sec.~\ref{Sec:Met} provides details on our numerical methodology to calculate {\it ab initio} the enthalpy landscape and details on the algorithmic searches. Sec.~\ref{Sec:Res} presents the results of the thermodynamic phase diagram, a comparison between two DFT functionals, the pressure dependence and a benchmark against an all-electron basis set. Finally we search for peaks in the electronic density of states near the Fermi level that could explain the experimental results. 

%=============================================
\section{Methods}\label{Sec:Met}
%=============================================

The compositional and configurational space of C-S-H was explored at 250 GPa and 300 GPa using the minima-hopping method~\cite{mh, Amsler2010, Sicher2011, Roy2008}. Minima-hopping is an efficient method to find low energy minima on the potential energy surface. In contrast to numerous other algorithms such as simulated annealing~\cite{sim_annealing} or basin hopping~\cite{basin_hopping} it is not based on thermodynamic principles. The method uses a clever combination of molecular dynamics and geometry optimizations to explore the entire configuration space without visiting known parts of it many times.

Three different sections of compositions in the ternary phase diagram were selected based on: I) experimental input (central region), II) high hydrogen content regions that make it more likely to find a high-temperature superconductor and III) the region designed as "doping" for its similarity to the \ce{H3S} phase.

\begin{figure}[t]
\includegraphics{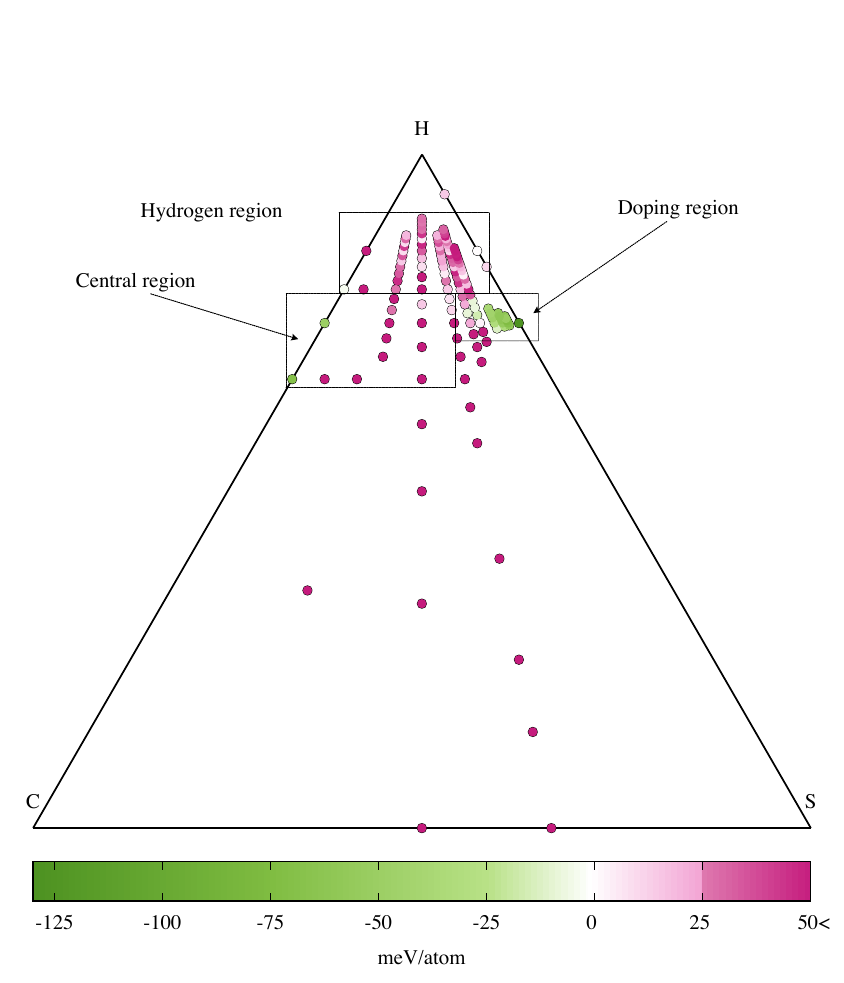} 
\caption{The ternary convex hull of formation enthalpy generated from 140 stochiometries explored at 250\,GPa. The formation enthalpy per atom can be read off the color bar. In addition we zoom into the three sections of interest that were studied in detail (central, high-H and doping) in Fig.~\labelcref{fig:tern_center,fig:tern_doping,fig:tern_hydro} respectively. }\label{fig:enth_250}
\end{figure} 

\begin{figure}[t]
    \includegraphics{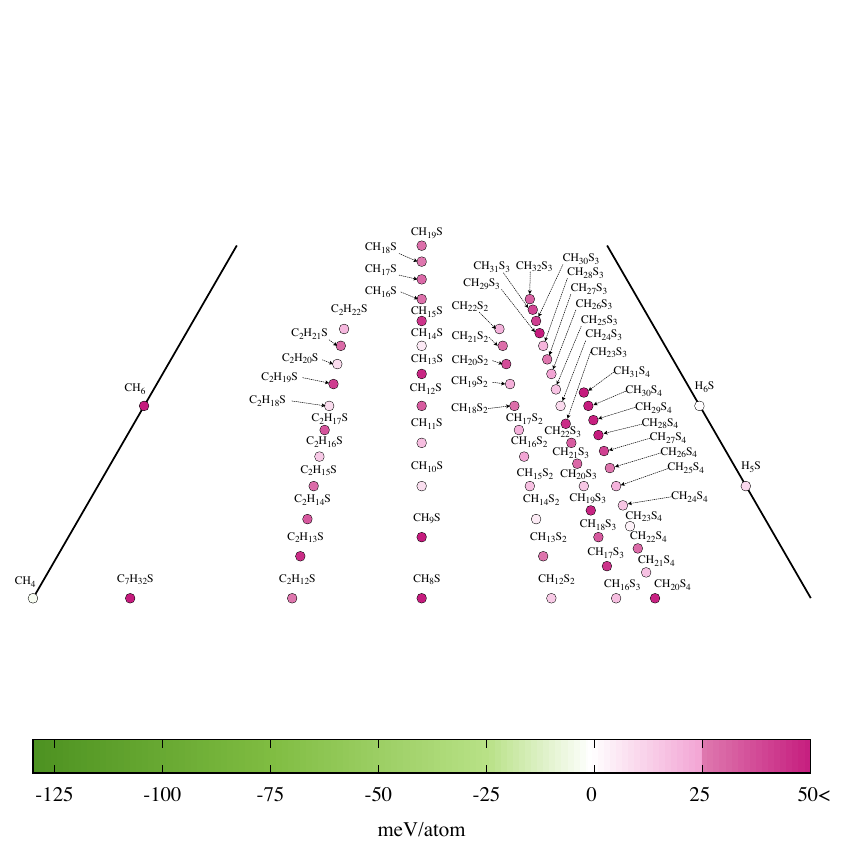} 
    \caption{Hydrogen rich region of the ternary phase diagram at 250 GPa}
    \label{fig:tern_hydro}
\end{figure}

\begin{figure}[t]
    \centering
    \includegraphics{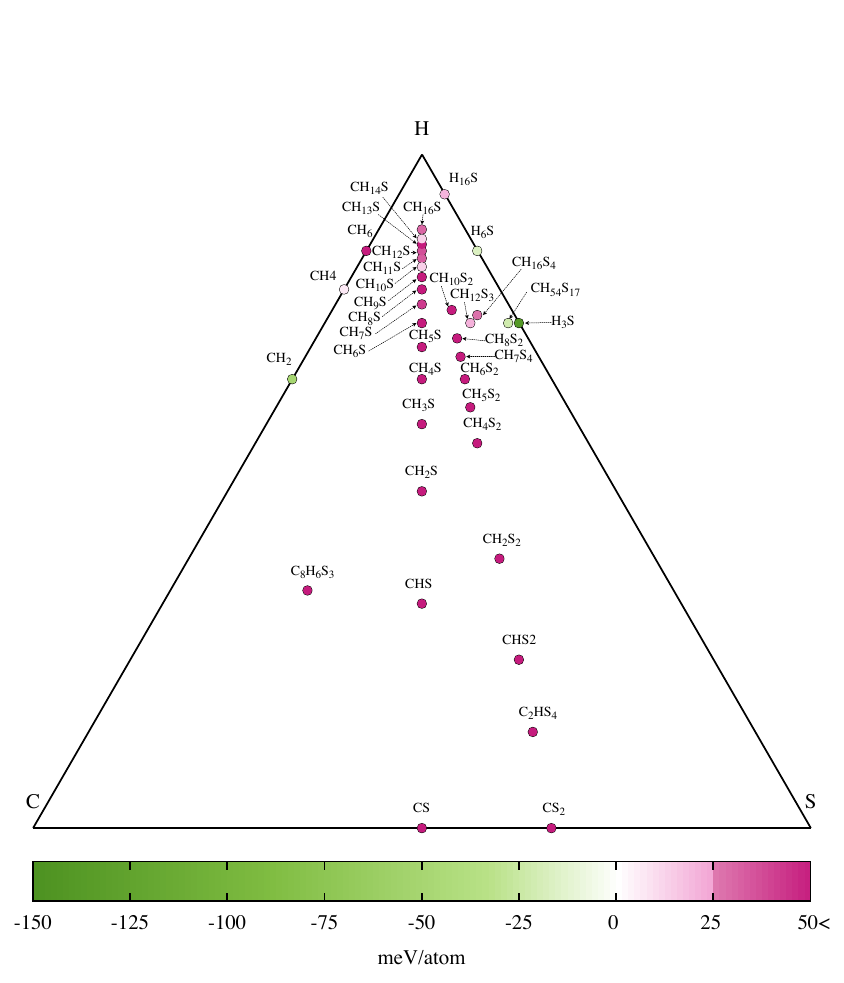}
    \caption{The ternary convex hull of formation enthalpy at 300 GPa}
    \label{fig:enth_300}
\end{figure}

We first scrutinized the 250\,GPa pressure range extensively and examined the 300\,GPa region afterwards. We invested the bigger part of our resources in the search at 250 GPa since this pressure range is reported most promising\cite{room_temp}.

Energy, atomic forces, and stresses were evaluated at the DFT level with the local density approximation (LDA) exchange-correlation functional. A plane wave basis-set with a cutoff energy of 300 eV was used to expand the wave function and in connection with the projector-augmented wave method. Geometry relaxations were performed with tight convergence criteria such that the forces on the atoms and the lattice derivatives were less than 5 meV/A.

Out of the more than 17,000 structures, 3000 were further studied. All structures that have an enthalpy close to the putative ground state (less than 20 meV per atom)
were selected for further analysis.

To eliminate possible inaccuracies from the pseudopotential at these high pressures, the selected systems were further analyzed with FHI-AIMS, an all electron electronic structure library. A full geometry relaxation (lattice vectors and atomic positions) was carried out with the tier 2 basis set of FHI-aims. Afterwards, the density of states was calculated and inspected for all these 3000 structures.

%=============================================
\section{Results}\label{Sec:Res} 
%=============================================

The first three subsections of this section discuss our findings at 250 GPa. The last subsection of Sec. \ref{Sec:Res} is dedicated to the results at 300 gigapascals.

\subsection{Structures and energies of the ternary phases at 250 GPa}

Fig.~\ref{fig:enth_250} shows the C-S-H ternary convex hull constructed for 140 compositions at 250 GPa. 
We subdivide it in three sections to improve readability and analyze regions of the phase diagram carefully. The first section is the central region, based on the experiments by Snider et al.~\cite{room_temp} and Lamichhane et al.~\cite{eos} in which carbon and sulfur were mixed at a 1:1 molar ratio and with a subsequent hydrogen incorporation. The second region is the high-hydrogen content section. The third section is the doping region, close to the \ce{H3S} superconductor. The colour bar at the bottom of the figure depicts the composition's enthalpy formation in meV/atom. 

The missing of a well defined ground state, whose energy is considerably lower than for related structures, is evident. Despite the use of larger simulation cells and substantially more extensive searches that found up to hundreds of structures per stochiometry, a decisive structural motif is missing in all structures. 
\begin{figure}[t]
    \includegraphics{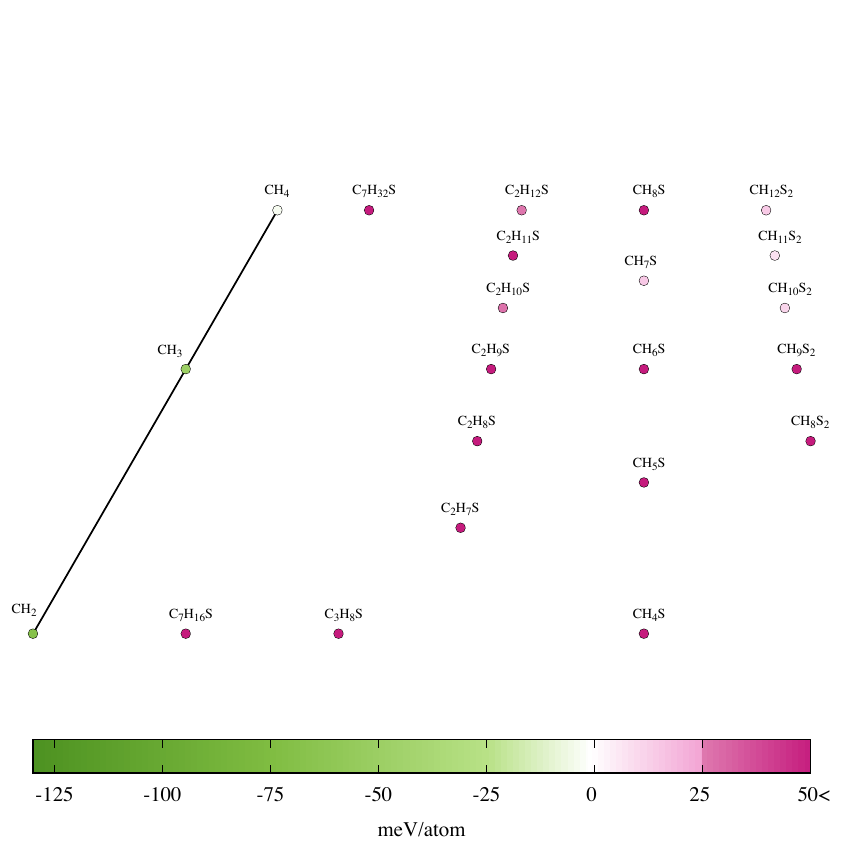}
    \caption{Center region of the ternary phase diagram at 250 GPa}
    \label{fig:tern_center}
\end{figure}

\begin{figure}[t]
    \includegraphics{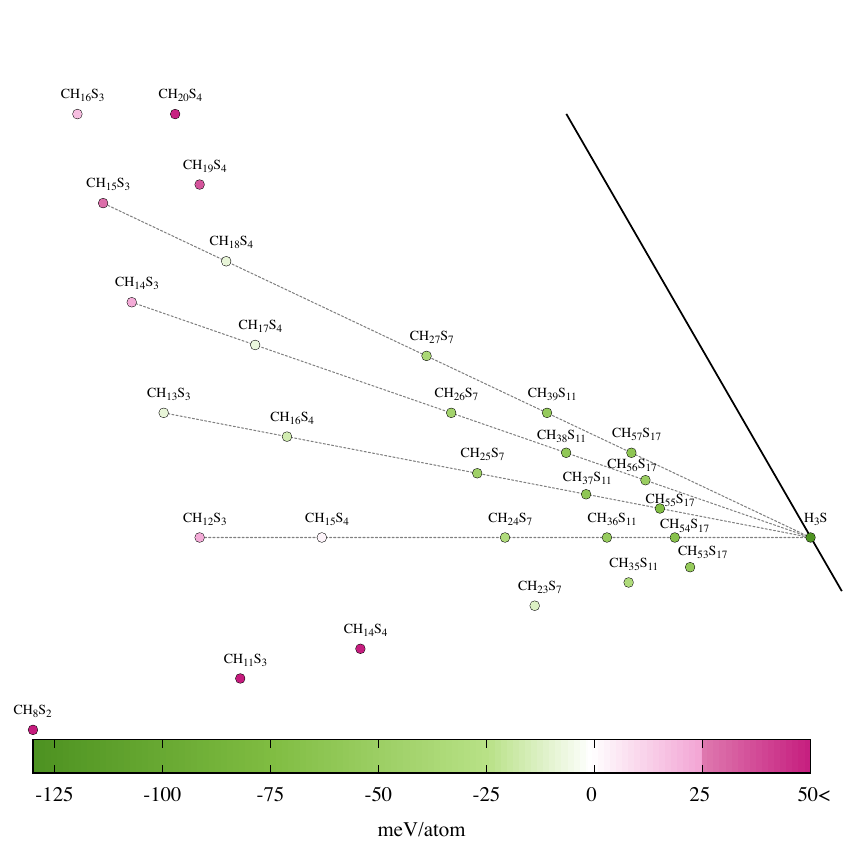}
    \caption{Carbon doping region of the ternary phase diagram at 250 GPa. The DOS of the structures lying on a grey dashed line are displayed in one of the plots of \figref{fig:doping}.}
    \label{fig:tern_doping}
\end{figure}

\begin{figure}[t]
    \includegraphics{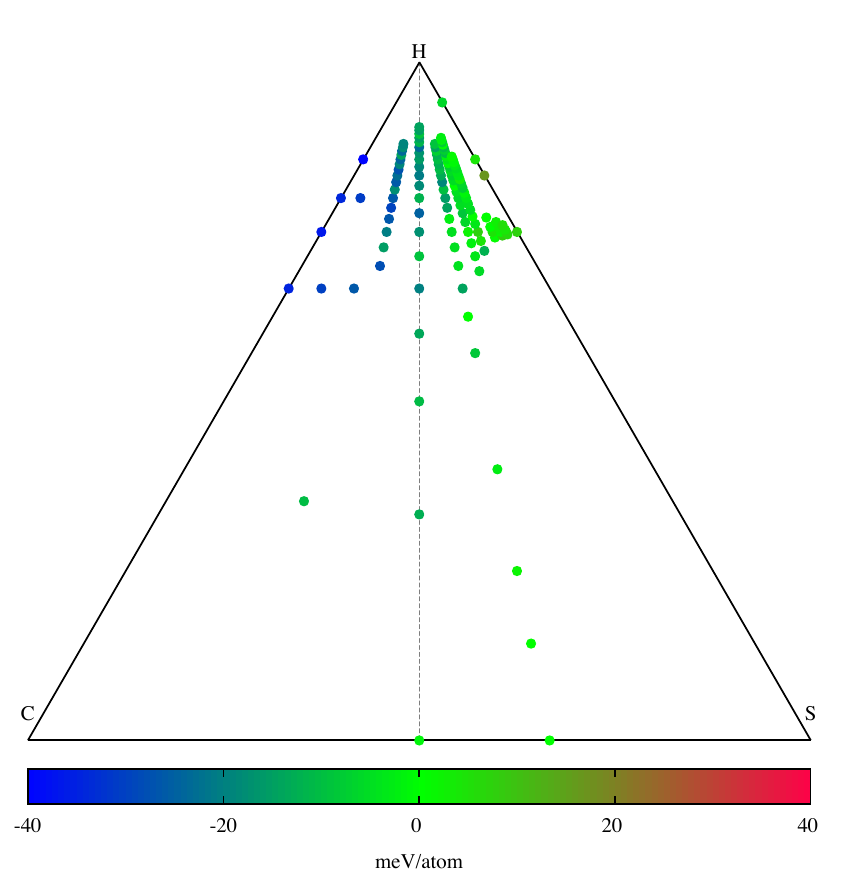}
    \caption{Comparison of formation enthalpy difference per atom between a PBE and an LDA functional at 250 GPa.}
    \label{fig:pbe_lda_compar}
\end{figure}
Let us start the analysis with the central area, depicted in detail in \figref{fig:tern_center}. According to the description of the experiments by Snider et al.; the RTS has likely been synthesized in this zone (at least at low pressure). Building blocks of CH$_2$, CH$_3$, and CH$_4$ are found in this region. The saturated carbon bonds lead to semiconducting phases with bandgaps between 0.7 and 1.2 eV. The possibility of sulfur doping CH$_x$ can be ruled out accordingly to \cite{doped_polyethylene}.

The hydrogen rich zone, shown in \figref{fig:tern_hydro}, was extensively reviewed in this work. In this region binaries appear as the lowest enthalpy structures, in this case, \ce{CH4} (-4~meV) followed by \ce{H6S} (-1~meV) and \ce{H5S} (9~meV). The best ternaries exhibit a slightly higher formation energy. The lowest ones are just a few meV per atom lower: \ce{CH10S} (8~meV), \ce{CH14S} (4~meV), \ce{CH14S2} (4~meV) and \ce{CH23S4} (2~meV). 
%We did not find any structure or main pattern through the enthalpies or the structures. 

\begin{figure*}[t]
\begin{subfigure}{.3\textwidth}
      \includegraphics[width=.9\linewidth]{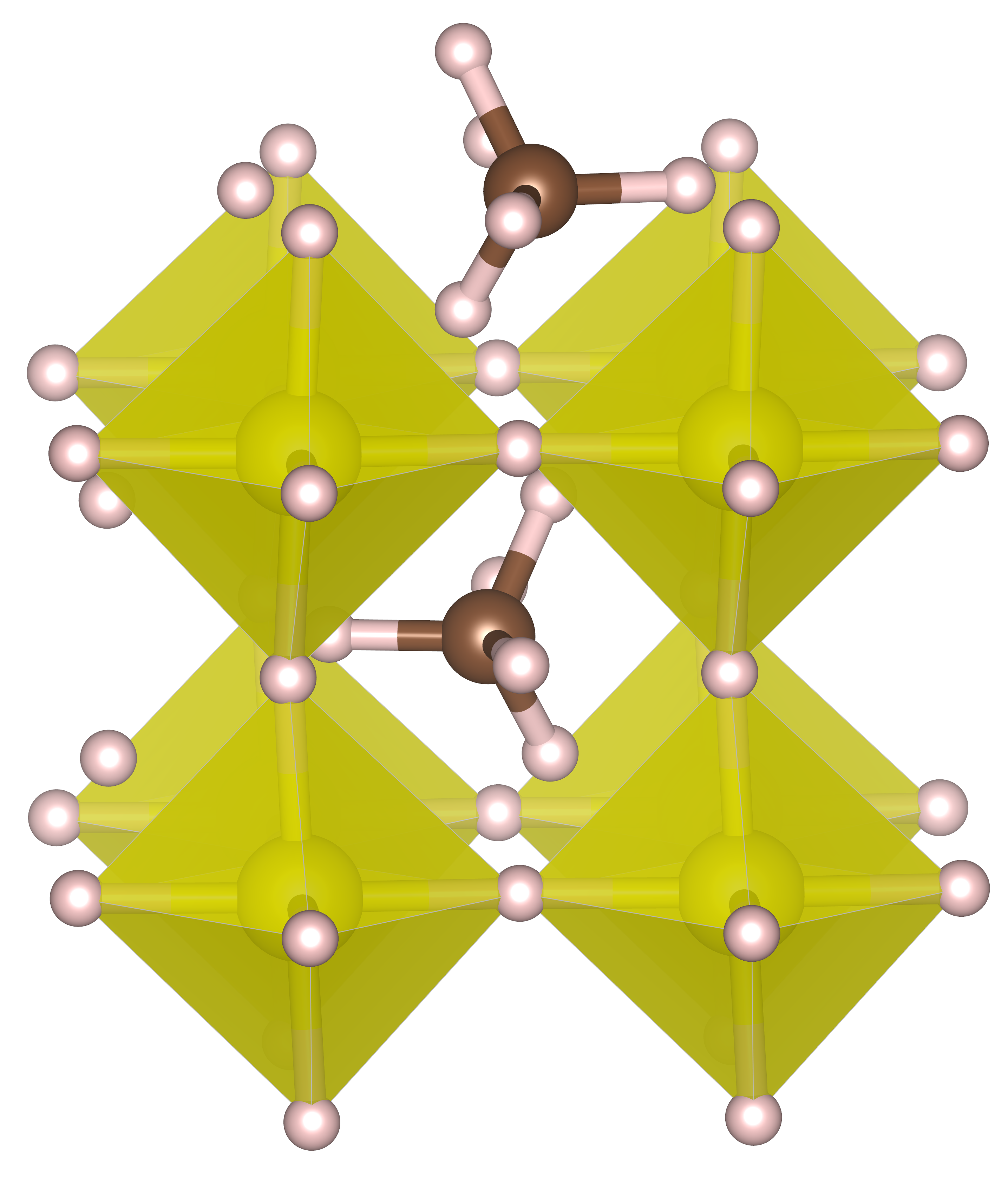}
    \caption{\ce{CH7S}}
    \label{fig:struc_ch7s}
\end{subfigure}%
\begin{subfigure}{.3\textwidth}
      \includegraphics[width=.9\linewidth]{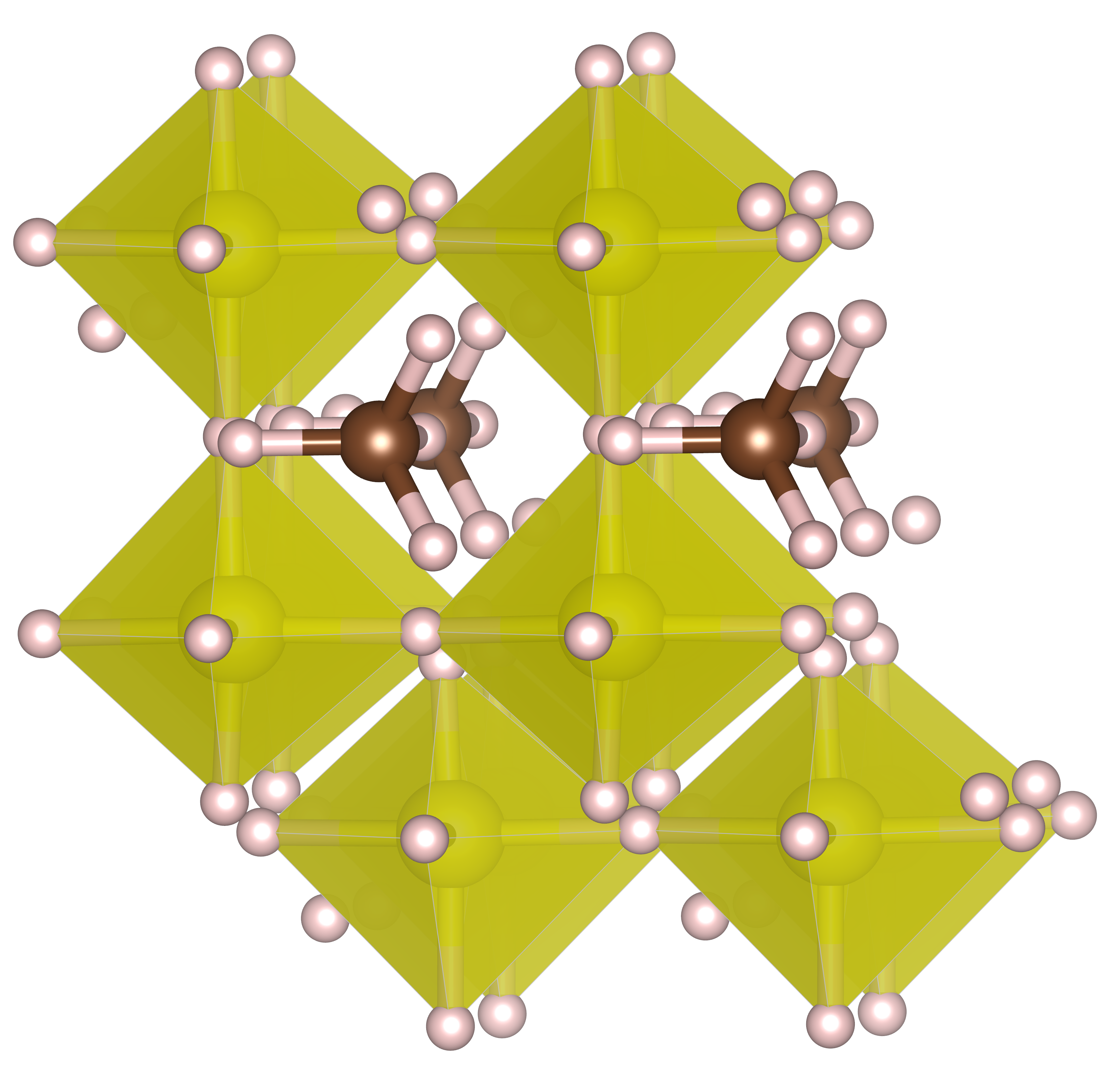}
    \caption{\ce{CH11S2}}
    \label{fig:struc_ch11s2}
\end{subfigure}%
\begin{subfigure}{.3\textwidth}
      \includegraphics[width=.9\linewidth]{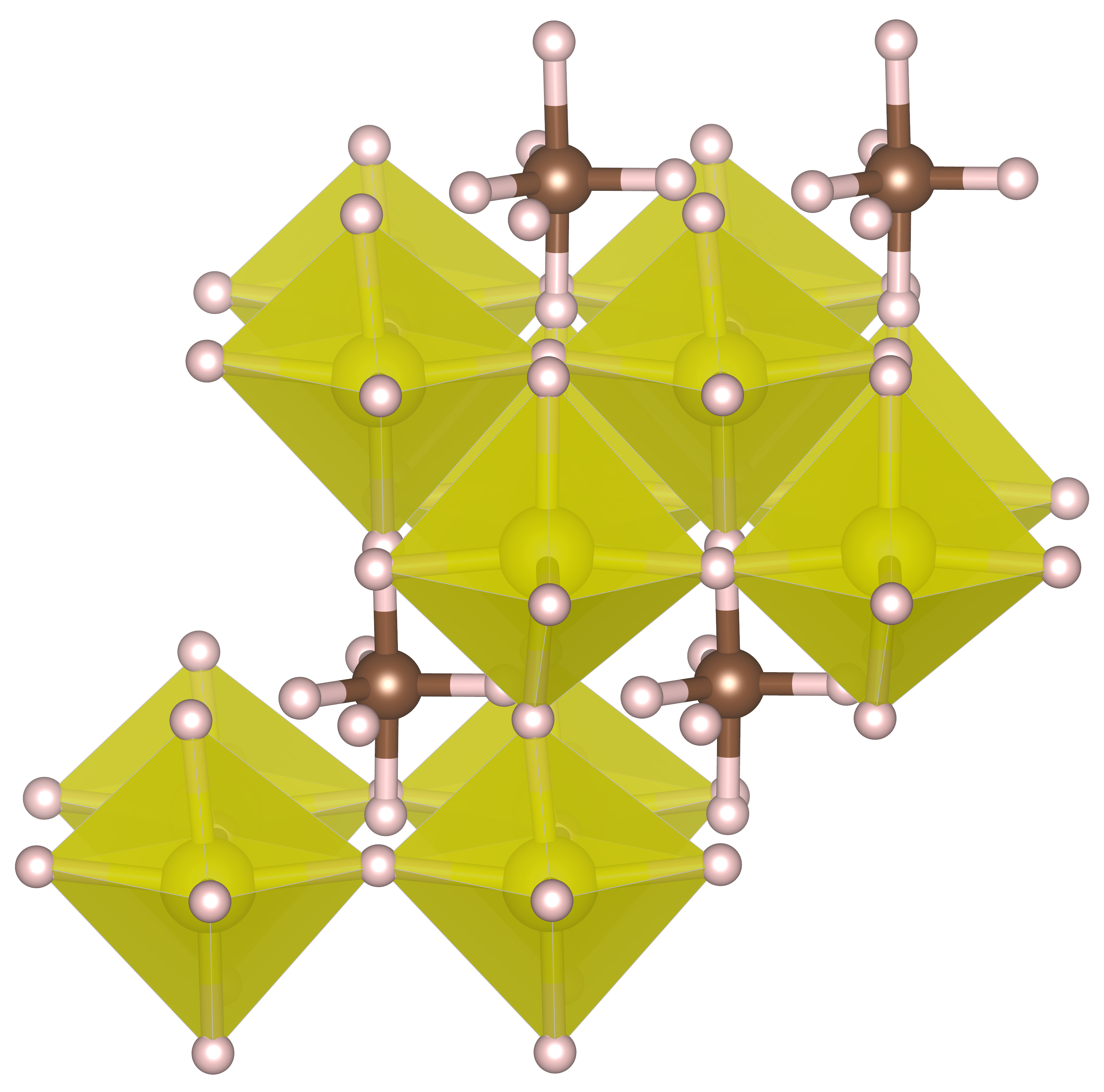}
    \caption{\ce{CH10S2}}
    \label{fig:struc_ch10s2}
\end{subfigure}
\begin{subfigure}{.3\textwidth}
      \includegraphics[width=.9\linewidth]{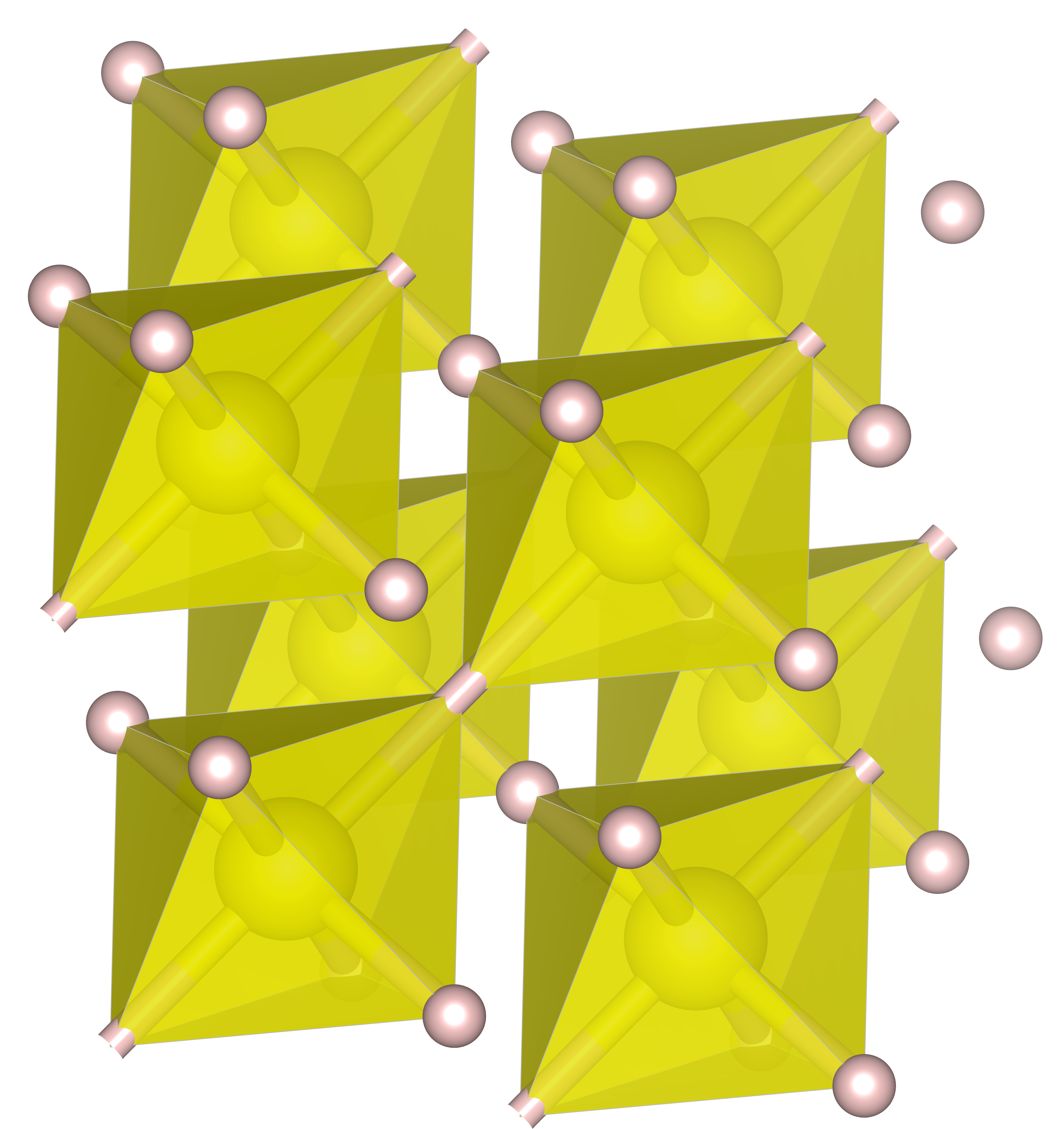}
    \caption{\ce{H3S}}
    \label{fig:struc_h3s}
\end{subfigure}%
\begin{subfigure}{.5\textwidth}
      \includegraphics[width=.9\linewidth]{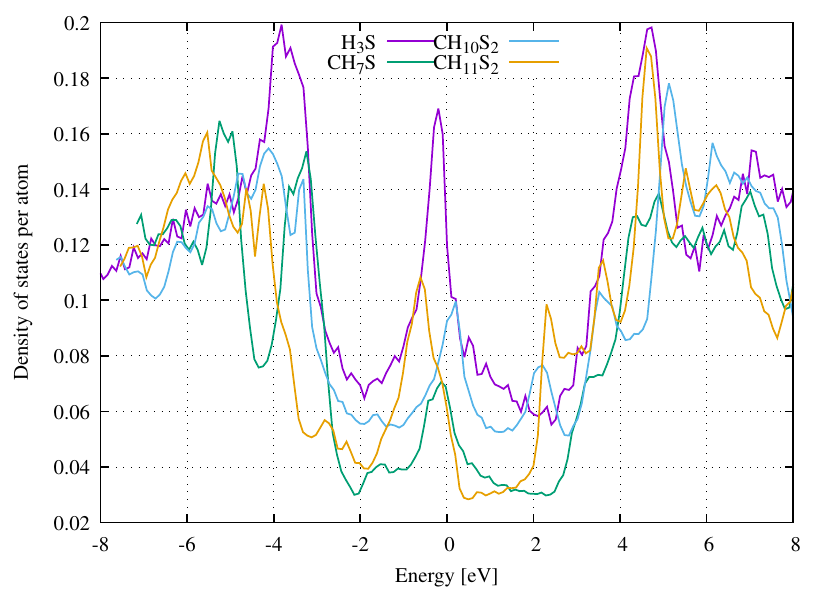}
    \caption{Density of states for \Cref{fig:struc_ch7s,fig:struc_ch11s2,fig:struc_ch10s2,fig:struc_h3s}}
    \label{fig:interesting_dos}
\end{subfigure}
\caption{A selection of the most promising structures with regards to their formation enthalpy and DOS at 250 GPa}
\label{fig:structures}
\end{figure*}

Finally, another region previously reported as promising is the doping area. In this work, we extended the search of the doping area conducted by Wang et al.\cite{Wang_absence_PRB-2021} by systematically varying carbon, sulfur and hydrogen as shown by the dashed lines (slopes) in \figref{fig:tern_doping}. The lowest energy doped system was 5.88\% carbon (\ce{CH53S17}). It is noticeable that the formation enthalpy decreases as one approaches the H3S phase. Also worth mentioning is the sudden jump of enthalpies, suggesting a regime change, going from \ce{CH_xS7} to \ce{CH_yS4}. \ce{H3S} is the most stable composition at 250 GPa with with a formation enthalpy of -130 meV. Its closest neighbours are carbon doped \ce{H3S} crystals which are the ternaries with the lowest formation enthalpy.

We also checked the influence of the exchange-correlation functional in describing the ternary's phase diagram. Fig.~\ref{fig:pbe_lda_compar} shows the enthalpy difference between LDA and GGA, at 250 GPa, for the lowest configuration for each stochiometry calculated by FHI-aims. Only minor differences are visible in the sulfur and hydrogen areas. Larger deviations start to appear towards the carbon areas of the ternary \ce{CH_xS} compositions lying on the middle line.

This can be explained by the well known over binding of LDA while PBE rather underbinds. The discrepancy between the two functionals is particularly pronounced for solids with molecular building blocks. Metallic systems are described reasonably well with both functionals. This is of course the most important class, since e do not expect that superconductivity arises from a semiconducting or insulating phase. 

\subsection{Low enthalpy compositions}

Among all the 140 studied compositions, the doping region appears as the most densely populated one. Other structures having a reasonably low formation enthalpy at 250\,GPa are \ce{CH7S}, \ce{CH10S2} and \ce{CH11S2}. Fig.~\ref{fig:structures} shows the crystalline structure for these compositions. On the left, the crystal model represents \ce{CH7S}, which belongs to the cubic crystal structure, pretty similar to the \cubic\ motif of \ce{H3S}. In this structure, every second sulfur atom and the corresponding hydrogen atoms are replaced by a \ce{CH4} molecule. The methane component changes its orientation along the z-axis. \ce{CH10S2} and \ce{CH11S2} (center and right models) in Fig.~\ref{fig:structures}, consist of two layers with \ce{H3S} structure separated by a carbon layer with different hydrogen environments.

\subsection{Electronic density of states}

\begin{figure*}[t]
  \begin{subfigure}{.45\textwidth}
    \includegraphics[width=.9\textwidth]{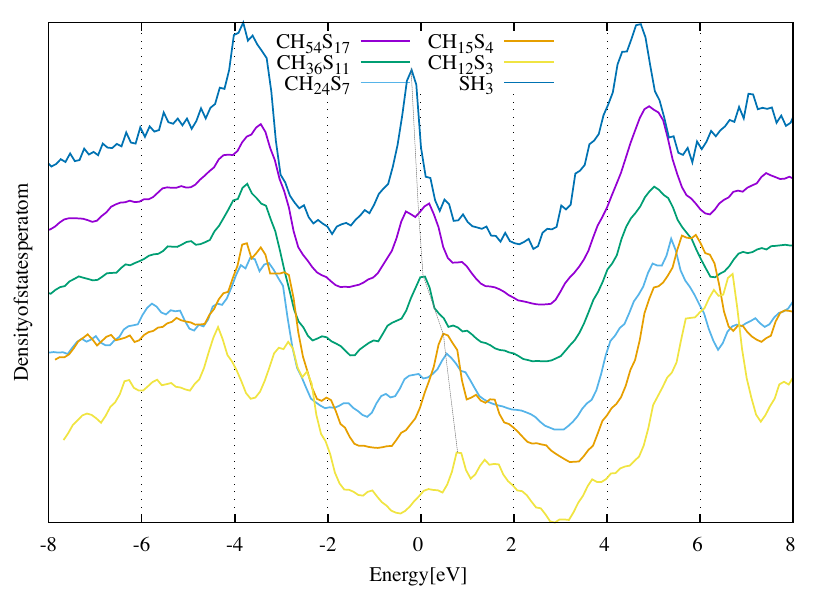}
    \caption{\ce{[CH3]_x[H3S]_{1-x}} \label{fig:dos3h}}
  \end{subfigure}%
  \begin{subfigure}{.45\textwidth}
    \includegraphics[width=.9\textwidth]{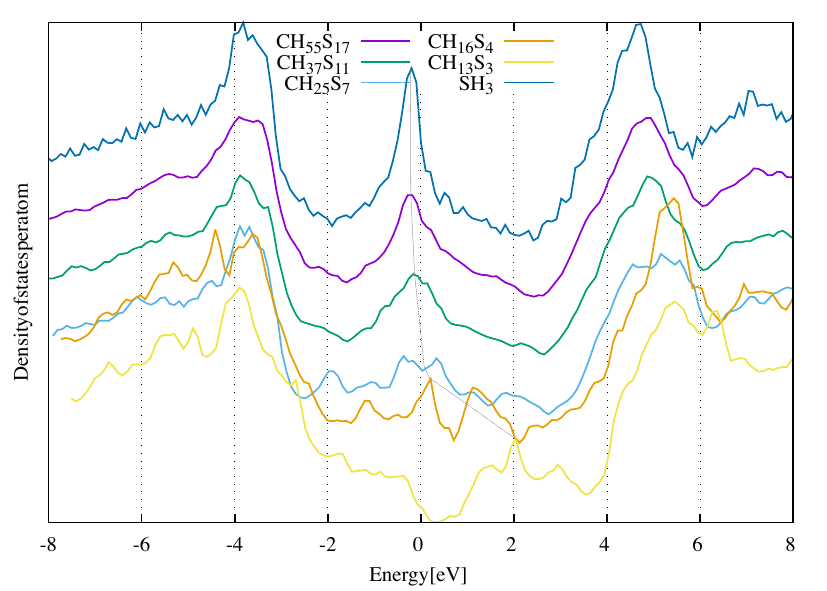}
    \caption{\ce{[CH4]_x[H3S]_{1-x}} \label{fig:dos4h}}
  \end{subfigure}
  \begin{subfigure}{.45\textwidth}
    \includegraphics[width=.9\textwidth]{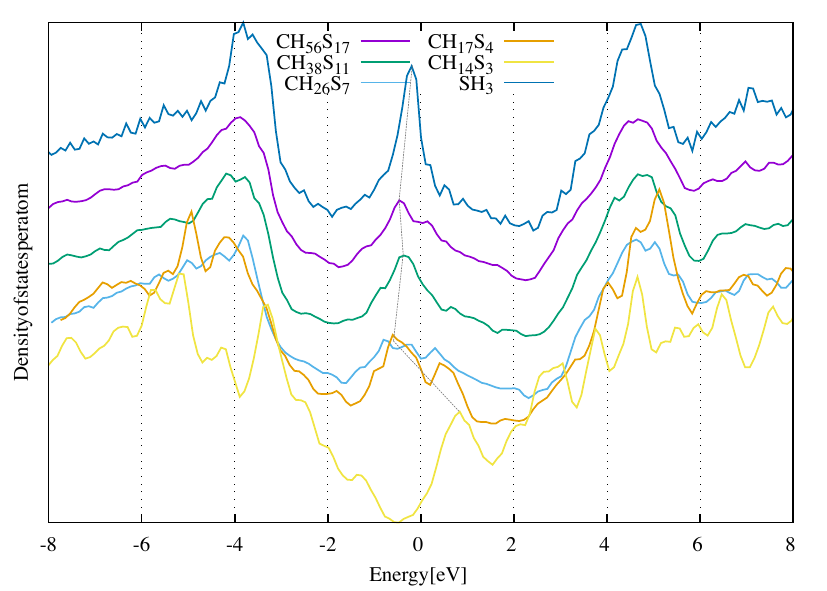}
    \caption{\ce{[CH5_x[H3S]_{1-x}} \label{fig:dos5h}}
  \end{subfigure}%
  \begin{subfigure}{.45\textwidth}
    \includegraphics[width=.9\textwidth]{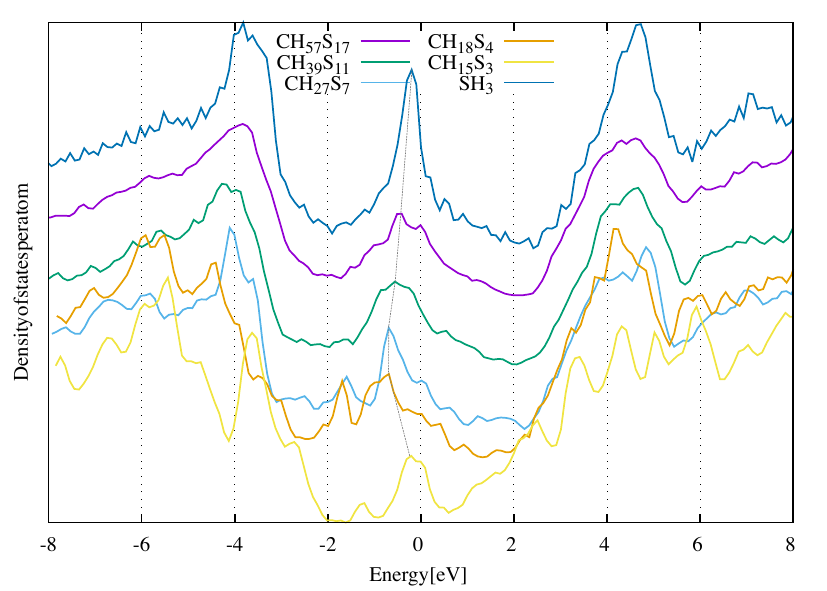}
    \caption{\ce{[CH6]_x[H3S]_{1-x}} \label{fig:dos6h}}
    \label{fig:sub2}
  \end{subfigure}
  \caption{Density of states in the carbon doping region at 250 GPa. The plots are normalized by the number of atoms and shifted downwards with increasing carbon doping ratio.}
  \label{fig:doping}
  \end{figure*}

The density of states was calculated and analyzed for the 3000 lowest local minima that were found. A selection of systems with the most promising DOS is discussed here. \figref{fig:doping} displays the DOS for most carbon doped \ce{H3S} structures. The lowering of the peak at the Fermi level with higher carbon concentration indicates inferior superconductive properties compared to \ce{H3S}. The peaks in DOS also move away from the Fermi level. In Figures \labelcref{fig:dos3h,fig:dos4h} they move to the right, in Figures \labelcref{fig:dos5h,fig:dos6h} they move to the left. This is indicated by the grey dashed lines in Figures \labelcref{fig:dos3h,fig:dos4h,fig:dos5h,fig:dos6h}. In the central region of the ternary phase diagram \ce{CH7S}, \ce{CH10S2} and \ce{CH11S2} have a low formation enthalpy and some interesting features in the DOS at the Fermi level displayed in \figref{fig:interesting_dos}. When their peak at the Fermi level is compared to the \ce{H3S} superconductor in \figref{fig:interesting_dos} none of them is as sharp as the one in the \ce{H3S} system which indicates a lower critical temperature than the one of \ce{H3S}. 
  
\subsection{Structures and energies of the ternary phases at 300 GPa}
\figref{fig:enth_300} displays the convex hull of formation enthalpy at 300 GPa. The structures assessed here are even less favourable against elemental decomposition than at 250 GPa. The only ternary which is favourable against elemental decomposition is \ce{CH54S17} with formation enthalpy of -22 meV per atom. This is a big increase in formation enthalpy compared to the one at 250 GPa where the formation enthalpy is -68 meV per atom.
Apart from a slight compression of the simulated cell, the \ce{CH54S17} structure is almost identical to the one at 250 GPa. \ce{H3S}, which is the most stable structure at 250 GPa. remains the most stable system at the higher pressure. In contrast to most other systems at 300 GPa its formation enthalpy per atom even drops from -130 meV per atom to -140 meV per atom while the structure remains unchanged. Some structures with a low formation enthalpy at 250 GPa had some kind of \ce{H3S} layers in it. An example of such structures can be seen in \figref{fig:structures}. At 300 GPa those layers can not be found anymore.

The DOS was also calculated and analyzed for all the structures at 300 GPa. No interesting peaks at the Fermi level were found.
  
%=============================================
\section{Conclusions}
%=============================================
A thermodynamically stable structure that gives rise to room temperature superconductivity is still elusive.
Of all the compositions and the two sets of pressures studied in this work, the lack of a broad, low-enthalpy formation region in the convex hull, similar to the one found for \ce{H3S}, poses more questions than it answers.
In the lack of further experimental evidence, we see the following possibilities: 

%scenario one
The right stochiometry of the superconductor was investigated in this and previous work, but the lowest enthalpy structure was not found. However, one would expect the RTS structure to be of negative formation enthalpy and its nearby composition and structures to yield a measurable negative formation enthalpy. At this stage and based on several previous and independent searches (using different searching algorithms), it is unlikely that an entire region of thermodynamically stable structures was overlooked.

%scenario two
Another possibility is that the structure lies in a part of the ternary phase diagram that was not explored. However, this also seems unlikely, considering our structure search covers most of the relevant regions of the phase diagram.

%scenario three
So it could be that Snider et al. have observed an unconventional room-temperature superconductor or another phenomenon with electrical resistance drops. In that case, all of the structures with sufficiently low formation enthalpy could be candidates for the RTS structure.

\section*{Structural Data}
A selection of the 3000 best structures and their density of states is available on this github repository: \url{https://github.com/moritzgubler/C-H-S_250GPa}. The atomic positions and the lattice constants were relaxed using the tier 2 basis of FHI-AIMS and a tight convergence criterion of $10^{-2}$ eV/A.

\begin{acknowledgments} 
The calculations were performed on the computational resources of the Swiss National Supercomputer (CSCS) under project s963 and on the Scicore computing center of the University of Basel.
\end{acknowledgments}
\newpage
\bibliographystyle{apsrev4-1}
\bibliography{main}

\end{document}